\documentclass{llncs}

\usepackage[utf8]{inputenc}
\usepackage[T1]{fontenc}

\usepackage{amssymb}
\usepackage{mathtools}
\usepackage{xfakebold}
\usepackage{xfp}
\newcommand{\indep}{\perp \!\!\! \perp}

\usepackage{booktabs}
\usepackage{makecell}

\usepackage{algorithm}
\usepackage{algorithmicx}
\usepackage[noend]{algpseudocode}
\algrenewcommand\textproc{}

\usepackage{graphicx}
\usepackage{subcaption}
\usepackage{float}

\usepackage{lipsum}

\usepackage{color}
\usepackage{soul}
\usepackage{mdframed}
\mdfsetup{backgroundcolor=yellow}

\usepackage{hyperref}
\hypersetup{
    colorlinks=true,
    linkcolor=blue,
    citecolor=blue,
    filecolor=blue,
    urlcolor=blue
}

\usepackage{cleveref}

\usepackage[backend=bibtex,style=ieee]{biblatex}
\usepackage{pifont}
\newcommand{\cmark}{\ding{51}}
\newcommand{\xmark}{\ding{55}}

\begin{document}

\title{
    The Impact of Missing Data on Causal Discovery: \newline
    A Multicentric Clinical Study
}

\author{
    Alessio Zanga \inst{1,2} \orcidID{0000-0003-4423-2121}
    \and Alice Bernasconi \inst{1,3} \orcidID{0000-0001-8522-6882}
    \and Peter J.F. Lucas \inst{4} \orcidID{0000-0001-5454-2428}
    \and Hanny Pijnenborg \inst{5} \orcidID{0000-0002-6138-1236}
    \and Casper Reijnen \inst{5} \orcidID{0000-0001-6873-7832}
    \and Marco Scutari \inst{6} \orcidID{0000-0002-2151-7266}
    \and Fabio Stella \inst{1} \orcidID{0000-0002-1394-0507}
}

\institute{
    Models and Algorithms for Data and Text Mining Laboratory (MADLab), University of Milano - Bicocca, Milan, Italy
    \and Data Science and Advanced Analytics, F. Hoffmann - La Roche Ltd, Basel, Switzerland
    \and Evaluative Epidemiology Unit, Department of Research, Fondazione IRCCS Istituto Nazionale dei Tumori, Milan, Italy
    \and University of Twente, Enschede, The Netherlands
    \and RadboudUMC, Nijmegen, The Netherlands
    Istituto Dalle Molle di Studi sull'Intelligenza Artificiale (IDSIA), Lugano, Switzerland
}

\maketitle

\section*{Abstract}
Causal inference for testing clinical hypotheses from observational data presents many difficulties because the underlying data-generating model and the associated causal graph are not usually available. Furthermore, observational data may contain missing values, which impact the recovery of the causal graph by causal discovery algorithms: a crucial issue often ignored in clinical studies. In this work, we use data from a multi-centric study on endometrial cancer to analyze the impact of different missingness mechanisms on the recovered causal graph. This is achieved by extending state-of-the-art causal discovery algorithms to exploit expert knowledge without sacrificing theoretical soundness. We validate the recovered graph with expert physicians, showing that our approach finds clinically-relevant solutions. Finally, we discuss the goodness of fit of our graph and its consistency from a clinical decision-making perspective using graphical separation to validate causal pathways.

\section{Introduction}
Much of the data collected in clinical research is observational in nature, that is, the data are collected as part of daily clinical practice.  As a consequence, interpreting the data correctly requires a good understanding of their characteristics and of the possible sources of bias. One of most well-known sources is missing values, which may arise in three different ways as categorized by Rubin \cite{Rubin1976InferenceData}:
\begin{itemize}
\item \emph{data missing completely at random} (MCAR), where the probability of a variable
being missing for a subject is independent of both observed and unobserved variables for that subject;
\item \emph{data missing at random} (MAR), where the probability of being missing for a subject is independent of the unobserved variables for that subject;
\item \emph{data missing not at random} (MNAR) that are neither MCAR nor MAR.
\end{itemize}
In the case of clinical observational data, MNAR is common and thus interesting to study, as it is often possible to unravel the reason for the missingness. The diagnostic process and the subsequent treatment selection are highly constrained by the condition of the patient, by the aim of limiting financial costs, and by minimizing the psychophysical burden for the patient. For instance, particular
laboratory tests may be skipped in favour of more precise ones that may be available at a later stage. This and similar processes produce missing values that are often systematic in nature.

In this paper, we focus on endometrial cancer (EC), a clinical domain where researchers try to unravel the causal mechanisms behind the prognosis of the disease, as is done in may other types of cancer. For the most part, existing literature is based on observational clinical data and assumes that they are MAR, while ignoring the \emph{likely} possibility that they may be MNAR. Our aim is to showcase how modern techniques from probabilistic graphical models can help dealing with the biases in causal discovery from observational data, in particular for MNAR. In order to explore the impact of incorrectly modeling missing data, we applied different causal discovery algorithms with different assumptions to data from a multicenter study on EC. In addition to discussing the statistical impact on the causal models, we highlight the clinical implications arising from the resulting biases.

\section{Related Research}

Missing values in clinical data are commonly imputed with heuristics like the ``hot deck'' and with single or multiple imputation. Such techniques assume that the data are MCAR or MAR: we cannot test whether this assumption is valid without knowing the missingness mechanism but, at the same time, if this assumption does not hold our clinical interpretations and conclusions are likely to be biased \cite{Stavseth2019HowData}. The most common scenario in clinical data, MNAR, is often ignored. Recently, we proposed a Causal Bayesian Network (CBN) approach \cite{Zanga2022RiskApproach} that extends a previous Bayesian network model for EC \cite{Reijnen2020PreoperativeStudy} designed to detect, estimate and control for potential selection bias. We highlighted the need for further investigation on the missing mechanism. Tu \emph{et al.} \cite{Tu2018CausalData} presented the MVPC algorithm for causal discovery with MNAR, extending the PC algorithm beyond the MAR setting. Similarly, in \cite{Liu2022GreedyValues} a new variant of the HC algorithm leveraged the definition of the missingness graph \cite{Mohan2018GraphicalData} to model the missingness mechanism.

\section{Background}
A causal graph $\mathcal{G} = (\mathbf{V}, \mathbf{E})$ \cite{pearl2016causal} is a directed acyclic graph (DAG) where for each directed edge $(X, Y) \in \mathbf{E}$, $X$ is a direct cause of $Y$ and $Y$ is a direct effect of $X$. The vertex set $\mathbf{V}$ is usually split into two disjoint subsets $\mathbf{V} = \mathbf{O} \cup \mathbf{U}$, where $\mathbf{O}$ is the set of the \emph{fully observed} variables, that is, variables with no missing values, while $\mathbf{U}$ is the set of \emph{fully unobserved} variables, also called the \emph{latent} variables. This partition allows us to express the values of an effect as a function of its direct causes: $V_i := f(Pa(V_i), U_i)$ where $Pa(V_i)$ is the set of variables with an edge into $V_i$. When missing values are present, causal graphs are not semantically adequate to express the missingness pattern and we have to resort to missingness graphs. A missingness graph $\mathcal{M} = (\mathbf{V}^*, \mathbf{E}^*)$ \cite{Mohan2018GraphicalData} is a causal graph where the vertices in $\mathbf{V}^*$ are partitioned into five disjoint subsets:
\begin{equation}
    \mathbf{V}^* = \mathbf{O} \cup \mathbf{U} \cup \mathbf{M} \cup \mathbf{S} \cup \mathbf{R}
\end{equation}
where $\mathbf{M}$ is the set of the \emph{partially observed} variables, that is, the variables with at least one missing value; $\mathbf{S}$ is the set of the proxy variables, that is, the variables that are actually observed; $\mathbf{R}$ is the set of the \emph{missingness indicators}:
\begin{equation}
    S_i := f(M_i, R_i) =
    \begin{cases}
        \; m_i & \text{if } r_i = 0, \\
        \; ?   & \text{if } r_i = 1.
    \end{cases}
\end{equation}
with $m_i$ the observed value of $M_i$ and `` ? '' a placeholder for the missing value. Since missingness graphs are causal graphs, they can be queried graphically for independencies between variables in the \emph{observed-data distribution} $P(\mathbf{O}, \mathbf{S}, \mathbf{R})$ by using \emph{d-separation}. The corresponding \emph{underlying} distribution with no missing values is $P(\mathbf{O}, \mathbf{M}, \mathbf{R})$. The set of variables $\mathbf{Z}$ d-separates $X$ from $Y$, denoted by $X \indep Y \mid \mathbf{Z}$, if it \emph{blocks} every path $\pi$ between $X$ and $Y$. A path $\pi$ is blocked by $\mathbf{Z}$ if and only if $\pi$ contains:
\begin{itemize}
    \item a fork $A \leftarrow B \rightarrow C$ or a chain $A \rightarrow B \rightarrow C$ so that $B$ is in $\mathbf{Z}$, or,
    \item a collider $A \rightarrow B \leftarrow C$ so that $B$, or any descendant of it, is not in $\mathbf{Z}$.
\end{itemize}
MCAR, MAR and MNAR result in different independence statements \cite{Rubin1976InferenceData} which can be assessed through d-separation. Therefore, we can link them to the independency statements they imply in the missingness graph:
    \begin{itemize}
        \item MCAR implies $\mathbf{O} \cup \mathbf{U} \cup \mathbf{M} \indep \mathbf{R}$: the missingness is random and independent from the fully observed and the partially observed variables.
        \item MAR implies $\mathbf{U} \cup \mathbf{M} \indep \mathbf{R} \mid \mathbf{O}$: missingness is random only conditionally on the fully observed variables;
        \item MNAR if neither MCAR nor MAR.
    \end{itemize}
Since MCAR implies MAR, any method assuming MAR can be used on MCAR.

\section{Causal Discovery from Observational Data}

When the causal graph $\mathcal{G}^*$ is unknown, it may be possible to recover it from the data $\mathcal{D}$ and the prior knowledge $\mathcal{K}$. This procedure is called \emph{causal discovery}; a recent survey is available from  \cite{Zanga2022APractice}. Clinical research relies mainly on observational data that often contains missing values, which may result in biased causal graphs depending on the missingness mechanism. This bias may lead causal discovery to include spurious associations in the graph \cite{Liu2022GreedyValues}. Clinicians usually perform multiple imputation which relies on the MAR assumption. Indeed, in our previous work on endometrial cancer \cite{Zanga2022RiskApproach}, we proposed a new causal discovery approach based on bootstrapping for clinical data with low sample size and high missing frequency assuming MAR, which we called the \emph{Bootstrap SEM}. While many methods to account for missing values in causal discovery under MCAR and MAR assumptions have been proposed in the literature, this was not the case for MNAR until recently when  algorithms such as \emph{MVPC} \cite{Tu2018CausalData} and \emph{HC-aIPW} \cite{Liu2022GreedyValues} were introduced. Both Bootstrap SEM and HC-aIPW are extensions of the \emph{Hill Climbing} (HC) algorithm that leverage missingness graphs. HC moves across the space of the possible causal graphs selecting the optimal graph $\mathcal{G}^*$ w.r.t. a goodness-of-fit function $\mathcal{S}$, known as the \textit{scoring criterion}. At its core, HC iteratively modifies the current recovered graph to maximize $\mathcal{S}$ by adding, deleting or reversing individual edges. When no modification improves the score, the procedure halts and returns the current solution. Bootstrap SEM nests the HC algorithm inside bootstrap resampling to mitigate the bias induced by the imputation of missing values via \emph{Structural Expectation-Maximization} (SEM) \cite{KollFried2009}. Finally, HC-aIPW performs an initial step to identify the parents set of the missingness indicators $\mathbf{R}$ via \emph{adaptive Inverse Probability Weighting} (aIPW) applied to the observed parent configurations, shaping the missingness mechanism of the variables $\mathbf{M}$. An important aspect that we must be aware of is that discovery algorithms are able to take advantage of the experts' knowledge $\mathcal{K}$ by including known causal statements. To allow for a fair comparison with Bootstrap SEM, we extend the HC-aIPW algorithm to encode this knowledge in forbidden and required edge lists, constraining the search space of candidate causal graphs.

\begin{table}[tbh]
  \caption{Variables in the dataset and their abbreviation; variables above
    the horizontal line are measured preoperatively, and those below the line
  postoperatively. Tier numbers are used as prior knowledge $\mathcal{K}$ in learning (see text).}
\label{tab-variables}
\vspace{0.5em}
\centering
\begin{tabular}{l|l|c}
\hline
\multicolumn{1}{c|}{\textbf{Variable}}
& \multicolumn{1}{c|}{\textbf{Abbreviation}}
& \multicolumn{1}{c}{\textbf{Tier}} \\
\hline\hline
Gynecological clinic (10 in total across Europe) & Hospital & 1 \\
Preoperative cervical cytology               & Cytology & 1 \\
Preoperative tumour grade                     & PreoperativeGrade & 0  \\
Cancer Antigen 125 serum levels              & CA125 & 1 \\
CT or MRI diagnostic imaging                 & CTMRI & 1 \\           
Estrogen receptor levels                     & ER & 1 \\              
Progesterone receptor levels                  & PR & 1 \\              
L1 cell adhesion molecule (cell motility)         & L1CAM & 1 \\
p53  (a tumour suppressor gene)               & p53 & 1 \\             
Platelets (number of platelets in blood)     & Platelets & 1 \\
\hline
Postoperative tumour grade                   & PostoperativeGrade & 2  \\
Lymphovascular space invasion                & LVSI & 2  \\
(Abdominal) lymph node metastases            & LNM & Free \\
Tumour invasion of myometrium                 & MyometrialInvasion & 2 \\
Treatment by chemotherapeutic drugs          & Chemotherapy & 2 \\
Treatment by radiation                       & Radiotherapy & 2 \\
Recurrence of the tumour                      & Recurrence & 3 \\           
Specific disease survival of at least $i$ years  & Survival$i$yr, $i \in \{1,3,5\}$ & 4 \\
\hline
\end{tabular}
\end{table}

\section{Multicentric Clinical Data on Endometrial Cancer}
\label{multicdata}

The observational data we explore in this paper
comprise 763 patients with endometrial cancer from 10
gynecological oncological clinics in Europe that are part of the
European Network for Individualized Treatment of Endometrial Cancer
(ENITEC). Clinical experts selected the variables 
that they considered most important for predicting the 
presence of lymph node metastases (LNMs) and survival
\cite{Reijnen2020PreoperativeStudy}. Approximately 10\% of endometrial
cancer patients present lymph node metastasis at diagnosis. The
selected variables, reviewed in Table \ref{tab-variables}, were
collected at the 10 different gynecological clinics where the patients were treated: the cytology of the cervix uteri, the preoperative tumour grade,
the postoperative tumour grade (after pathological examination of the
tumour tissue obtained after surgical removal of the uterus), treatment
by chemotherapy or radiatiotherapy, lymphvascular space invasion (that is,
whether there is tumour growth into the lymph or blood vessels), the
levels of estrogen and progesterone in blood, the presence of lymph
node metastasis according to CT or MRI imaging, the CA125 tumour marker,
L1CAM (an intracellular protein that promotes tumour cell
motility), the p53 tumour suppressor gene, the number of platelets,
presence of lymph node metastases, recurrence of the tumour, and
lastly survival before and after 1, 3, and 5 years. The tumour
markers, such as p53, CA125, L1CAM, estrogen and progesterone levels
are thought to offer causal prognostic information about tumour cell
behaviour and thus tumour ingrowth, metastases, recurrence, and
survival. Causal discovery taking into account sources
of biases may offer new insight into how these variables interact.

\section{Experiments}
We performed numerical experiments to compare the graphs recovered under MAR and MNAR by the Bootstrap SEM and HC-aIPW algorithms respectively. Data and code are available upon request by following the procedure described in \cite{Reijnen2020PreoperativeStudy}. For reference, we also reported the results for HC on data completed performing single imputation, which we refer to as \emph{HC-complete}. Prior knowledge $\mathcal{K}$, elicited by experts,  consists of forbidden edges, variables in higher tiers (\Cref{tab-variables}) are not allowed to cause those in lower tiers, and required edges, Survival$1$yr is a direct cause of Survival$3$yr, and Survival$3$yr is a direct cause of Survival$5$yr.

\subsection{Quantitative Analysis in Terms of Goodness-of-Fit}

\begin{figure}[b!]
    \vspace{-0.50cm}
    \centering
    \includegraphics[scale=0.80]{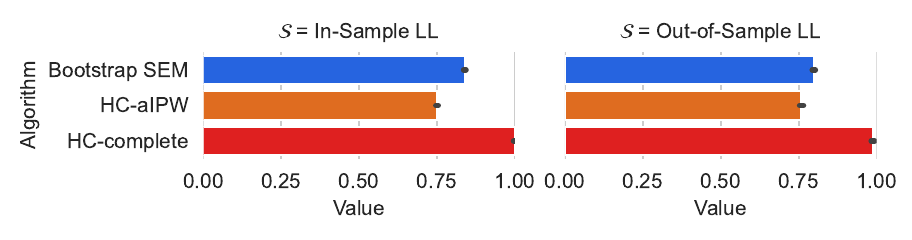}
    \caption{In-sample and out-of-sample LL for each algorithm. Re-scaled by sample size and absolute maximum value. Lower values are better.}
    \label{fig:goodness_of_fit}
    \vspace{-0.50cm}
\end{figure}

Firstly, we evaluated the goodness of fit of the recovered graphs by computing the log-likelihood (LL) of both the data used to recover the graph (in-sample) as well as that of data held aside for validation (out-of-sample). The former allows us to see which algorithm fits a particular data set the best; the latter approximates the Kullback-Leibler distance between the recovered causal graph and the unknown causal graph underlying the data, and allows us to see how close the two are in the space of the possible causal graphs. It also provides a measure of how well the recovered graph generalises to new data \cite{KollFried2009}.
We repeated causal discovery for 100 bootstrap replicates and computed mean and standard deviation of both in-sample and out-of-sample LL to improve accuracy and assess their variability. The results are reported in \Cref{fig:goodness_of_fit}.


We observe that HC-aIPW, which assumes MNAR, dominates Bootstrap SEM and HC-complete, which assume MAR, for both in-sample and out-of-sample LL. In the case of in-sample LL, this may be attributed (at least in part) to making the correct assumption about the missingness mechanism: assuming MCAR or MAR would be too restrictive and thus limit how well the recovered graph can fit the data. On the other hand, assuming MNAR when MCAR or MAR would be suitable is likely to let causal discovery algorithms overfit the model because it is not strict enough as an assumption. This would decrease the out-of-sample LL because overfitted models do not generalize well and are much more complex than the model underlying the data. The fact that HC-aIPW, which assumes MNAR, outperforms Bootstrap SEM and HC-complete, which assume MAR, in terms of in-sample LL while also outperforming them in out-of-sample LL suggests that the MNAR assumption is correct for the data, and that assuming MNAR allows HC-aIPW to recover a causal graph that is close to the model underlying the data and to generalise better to new data as a result. 
We also observe that HC-complete is the worst among the considered algorithms: it is also the only one that incorporates missing data before learning, instead of learning the missingness mechanisms along with the causal graph. This suggests that the latter approach can produce better results. Finally, we observe a smaller reduction from in-sample LL to out-of-sample LL in HC-aIPW (ratio: 0.43) than in Bootstrap SEM (ratio: 0.40). Larger reductions may be indicative of overfitting (i.e. fitting the data too well at the cost of losing the ability to generalise), which suggests that HC-aIPW might be less prone to overfitting.

\subsection{Qualitative Analysis in Terms of Independencies}

\begin{figure}[H]
    \vspace{-1.50cm}
    \centering
    \includegraphics[scale=0.7]{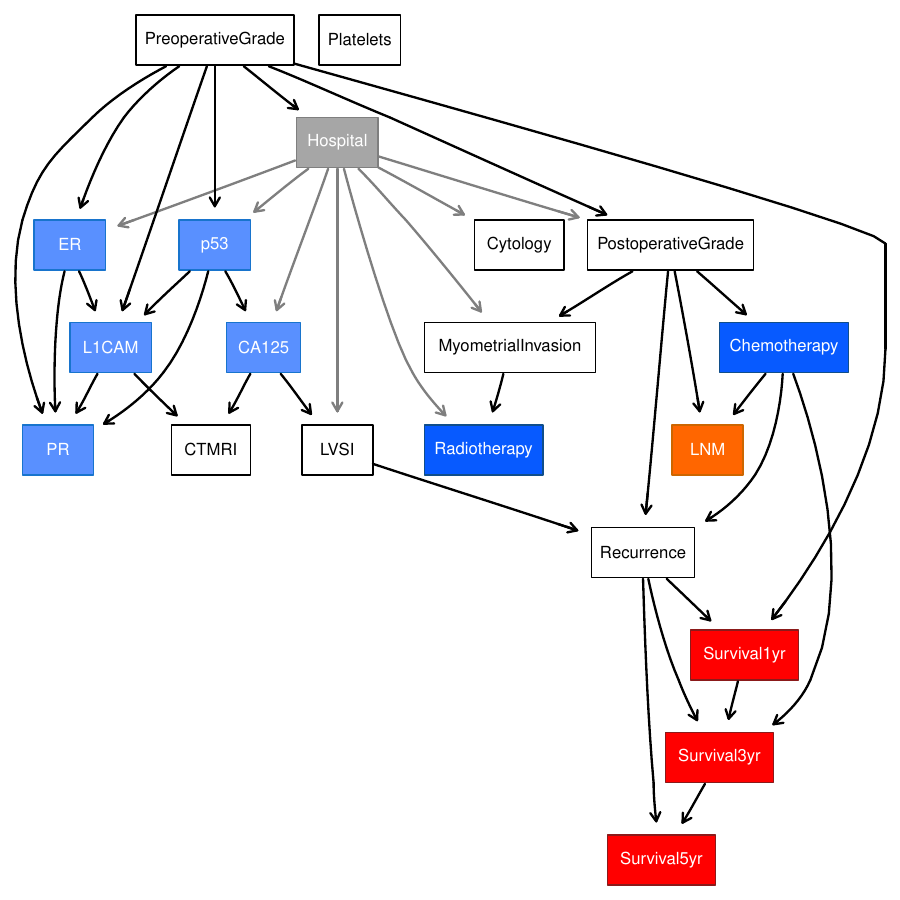}
    \caption{Causal graph $\mathcal{M}_{\textrm{MAR}}$ recovered by Bootstrap SEM under MAR.
    }
    \label{fig:hc_bootstrap_dag}

    
    \includegraphics[scale=0.7]{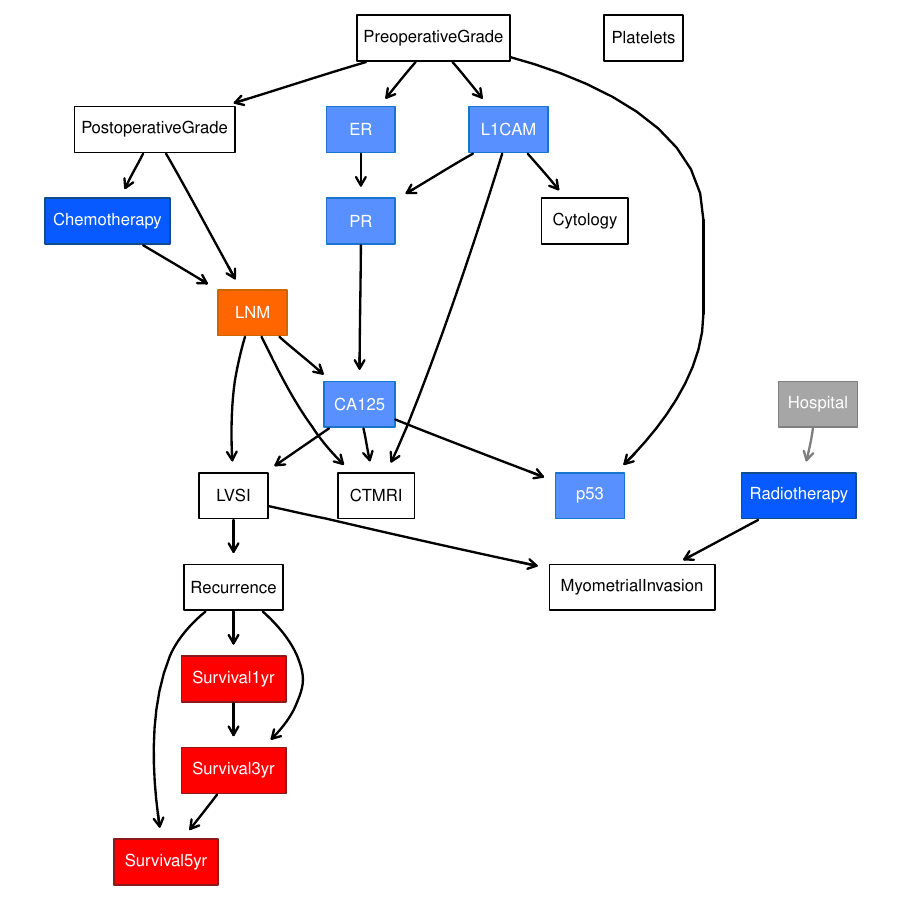}
    \caption{Causal graph $\mathcal{M}_{\textrm{MNAR}}$ recovered by HC-aIPW under MNAR.
    }
    \label{fig:hc_missing_dag}
\end{figure}

While goodness-of-fit measures are capable to provide a \emph{quantitative} evaluation of the recovered graphs, they say little about the \emph{qualitative} evaluation of the information encoded in such graphs. \Cref{fig:hc_bootstrap_dag} is a graphical representation of the graph recovered by Bootstrap SEM,  $\mathcal{M}_{\textrm{MAR}}$, while \Cref{fig:hc_missing_dag} is shows that recovered by HC-aIPW, $\mathcal{M}_{\textrm{MNAR}}$. For readability, we colored the vertices depending on their \emph{semantic} interpretation. Therefore, the treatments (i.e. Radiotherapy and Chemotherapy) are colored in blue, the outcomes (i.e. Survival1yr, etc.) are red, the event of interest (i.e. LNM) is orange and the relevant biomarkers (i.e. ER, PR, CA125, etc.) are lightblue. Finally, the \emph{context} variable, or selection variable, Hospital is colored in gray.

A complementary evaluation to scoring criteria can be done in terms of \emph{independence statements} using d-separation. We report in \Cref{tab:d_sep} the consistency evaluation of our findings w.r.t. clinical knowledge on EC. Focusing on the interactions of LNM, we observe that $\textrm{LNM} \indep \{\textrm{CA125}, \textrm{p53} \} \mid \textrm{PostoperativeGrade}$ is true in $\mathcal{M}_{\textrm{MAR}}$, but false in $\mathcal{M}_{\textrm{MNAR}}$, where CA125 and p53 are effects of LNM. This evidence makes $\mathcal{M}_{\textrm{MNAR}}$ close to the clinical practice where both CA125 and p53 are considered relevant biomarkers linked to LNM, providing additional information on LNM even when PostoperativeGrade is observed. If LNM were a \emph{leaf} (that is, a vertex without outgoing edges), as in $\mathcal{M}_{\textrm{MAR}}$, this would mean that it is not a cause of any other variable, which is clearly a contradiction: how could LNM not be the cause, neither directly nor indirectly, of the level of a biomarker that, according to physicians, is supposed to be connected with? Indeed, one of the crucial differences between $\mathcal{M}_{\textrm{MAR}}$ and $\mathcal{M}_{\textrm{MNAR}}$ is the interaction between the biomarkers and LNM: while they are d-separated in the former graph when PostoperativeGrade is observed, biomarkers are descendants of LNM in the latter. Hence, if our goal is to detect the presence of LNM in EC patients, then measuring CA125, p53 or any of their descendants is coherent with the statements encoded into $\mathcal{M}_{\textrm{MNAR}}$.

\begin{table}[b!]
    \centering
    \vspace{-0.75cm}
    \caption{D-separations that do (\cmark) or do not hold (\xmark) in $\mathcal{M}_{\textrm{MAR}}$ and $\mathcal{M}_{\textrm{MNAR}}$.}
    \label{tab:d_sep}
    \vspace{0.5em}
    \begin{tabular}{l|cc}
        \hline
        \textbf{Statement} &
        $\quad\pmb{\mathcal{M}_{\textrm{MAR}}}\quad$ &
        $\pmb{\mathcal{M}_{\textrm{MNAR}}}\quad$ \\
        \hline
        $\textrm{LNM} \indep \{\textrm{CA125}, p53 \} \mid  \textrm{PostperativeGrade} \quad$   & \cmark & \xmark \\
        $\textrm{LNM} \indep \textrm{Chemotherapy}$                   & \xmark & \xmark \\
        $\textrm{LNM} \indep \textrm{Radiotherapy}$                   & \xmark & \cmark \\
        $\textrm{LNM} \indep \textrm{Hospital}$                   & \xmark & \cmark \\
        \hline
    \end{tabular}
    \vspace{-1.50cm}
\end{table}

Shifting the focus to the treatment variables Chemotherapy and Radiotherapy, $\textrm{LNM} \indep \textrm{Chemotherapy}$ does not hold either in $\mathcal{M}_{\textrm{MAR}}$ or in $\mathcal{M}_{\textrm{MNAR}}$, since Chemotherapy is a direct cause of LNM in both causal graphs. This means that Chemotherapy is expected to influence the likelihood of LNM, which is exactly the reason why it is prescribed by clinicians. On the other hand, $\textrm{LNM} \indep \textrm{Radiotherapy}$ does hold in  $\mathcal{M}_{\textrm{MNAR}}$, but not in $\mathcal{M}_{\textrm{MAR}}$, suggesting that the observed spurious correlation related to the decision of performing Radiotherapy is induced by an MNAR pattern. This is confirmed by the clinical literature: since Radiotherapy is aimed at local treatment of the tissue surrounding the uterus, and there is a clear dependence with MyometrialInvasion of the tumour (in both models), Radiotherapy effects on LNM are not expected \cite{radiotherapy}.

Another major difference between $\mathcal{M}_{\textrm{MAR}}$ and $\mathcal{M}_{\textrm{MNAR}}$ is related to the the edges incident on the Hospital variable. While in $\mathcal{M}_{\textrm{MAR}}$ this variable is connected to biomarkers, treatments and others variables, in $\mathcal{M}_{\textrm{MNAR}}$ it is the only parent of Radiotherapy. This difference is represented by the statement $\textrm{LNM} \indep \textrm{Hospital}$, which is false in $\mathcal{M}_{\textrm{MAR}}$ but \emph{true} in $\mathcal{M}_{\textrm{MNAR}}$. It is crucial to underline that Hospital is a context variable, that is, it represents the context in which the data have been collected. Hence, the observed difference in the associated (unconditional) statement could lead to an erroneous evaluation of the impact of the selection bias. In particular, since this case study is a multicentric study on LNM in EC, observing a dependence between LNM and Hospital, as in $\mathcal{M}_{\textrm{MAR}}$, suggests that there are unobserved factors associated to the environment that affect LNM. This is clearly not the case in $\mathcal{M}_{\textrm{MNAR}}$, where LNM is independent of Hospital and becomes dependent only by conditioning on MyometrialInvasion. In fact, MyometrialInvasion is the only common descendant of LNM and Hospital. Because gynecologists involved in collecting the data are
collaborating in ENITEC (See Section \ref{multicdata}), it can be expected that
data will be very similar because of the use of guidelines in treatment.
However, Radiotherapy is prescribed in hospitals with different likelihoods, which
is reflected in particular in $\mathcal{M}_{\textrm{MNAR}}$.

\section{Discussion \& Conclusions}

In this work we presented a systematic analysis of the impact of missingness assumptions using state-of-the-art causal discovery algorithms. We applied these methods to real-world observational data collected in a multicentric study on endometrial cancer patients, addressing the issues of low sample size and high missing value frequency. We extended existing algorithms to include experts' prior knowledge, without sacrificing theoretical soundness. Furthermore, we validated the obtained causal models with experienced physicians, checking their consistency against current clinical literature. We evaluated the goodness-of-fit of the recovered graphs with respect to the underlying data distribution, showing that stricter assumptions are associated to models that generalize poorly. Moreover, by leveraging the test for graphical separation, we explained how the missingness mechanism affects the causal pathways associated to the clinical decision-making perspective. While causal discovery in presence of missing values shed light on the effect of missingness bias in clinical decision making, the problem is far from being solved. For example, it would be crucial to assess the size of the distortion and its impact in others specific case of study. Finally, it would be interesting to analyze settings in which the missingness bias is overlapped with hidden variables and selection variables, disentangling the interaction of cause-effect pairs in observational studies.

\section*{Acknowledgments}
Alessio Zanga is funded by F. Hoffmann-La Roche Ltd.

\printbibliography

\end{document}